\documentclass[aps,prl,reprint,preprintnumbers,superscriptaddress,amsmath,amssymb,bibnotes,longbibliography]{revtex4-1}

\usepackage{graphicx}
\usepackage{dcolumn}
\usepackage{bm}
\usepackage{color}
\usepackage[colorlinks,linkcolor=blue,anchorcolor=blue,citecolor=blue,breaklinks,CJKbookmarks=True,urlcolor=blue,filecolor=blue,menucolor=blue,runcolor=blue]{hyperref}

\begin{document}


\title{Superconductivity of cerium under pressures up to 54GPa}

\author{Y. N. Zhang}
\thanks{These authors contributed equally to this work.}
\affiliation  {Center for Correlated Matter and School of Physics, Zhejiang University, Hangzhou 310058, China}
\author{D. J. Su} 
\thanks{These authors contributed equally to this work.}
\affiliation  {Center for Correlated Matter and School of Physics, Zhejiang University, Hangzhou 310058, China}

\author{Z. Y. Shan}
\affiliation  {Center for Correlated Matter and School of Physics, Zhejiang University, Hangzhou 310058, China}

\author{Z. H. Yang}
\affiliation  {Center for Correlated Matter and School of Physics, Zhejiang University, Hangzhou 310058, China}

\author{J. W. Zhang}
\affiliation  {Center for Correlated Matter and School of Physics, Zhejiang University, Hangzhou 310058, China}

\author{R. Li}
\affiliation  {Center for Correlated Matter and School of Physics, Zhejiang University, Hangzhou 310058, China}

\author{M. Smidman}
\affiliation  {Center for Correlated Matter and School of Physics, Zhejiang University, Hangzhou 310058, China}

\author{H. Q. Yuan}
\email[Corresponding author: ]{hqyuan@zju.edu.cn}
\affiliation{Center for Correlated Matter and School of Physics, Zhejiang University, Hangzhou 310058, China}
\affiliation  {State Key Laboratory of Silicon and Advanced Semiconductor Materials, Zhejiang University, Hangzhou 310058, China}
\affiliation  {Collaborative Innovation Center of Advanced Microstructures, Nanjing 210093, China}

\date{\today}

\begin{abstract}
Cerium is a fascinating element due to its diverse physical properties, which include forming various crystal structures ($\gamma$, $\alpha$, $\alpha^{'}$, $\alpha^{''}$ and $\epsilon$), mixed valence behavior and superconductivity, making it an ideal platform for investigating the interplay between different electronic states. Here, we present a comprehensive transport study of cerium under hydrostatic pressures up to 54 GPa. Upon applying pressure, cerium undergoes the $\alpha$ $\rightarrow$ $\alpha^{''}$ transition at around 4.9~GPa, which is accompanied by the appearance of superconductivity with $T_{\rm c}$ of 0.4~K, and $T_{\rm c}$ slightly increases to 0.5~K at 11.4~GPa. At 14.3~GPa, $T_{\rm c}$ suddenly increases when the $\alpha^{''}$ phase transforms into the $\epsilon$ phase, reaching a maximum value of 1.25~K at around 17.2~GPa.  Upon further increasing the pressure, $T_{\rm c}$ monotonically decreases. Together with the results of previous studies, our findings suggest that the evolution of superconductivity in cerium is closely correlated with the multiple pressure-induced structural transitions and corresponding unusual electronic structures.
\end{abstract}
\maketitle


\section{\uppercase\expandafter{\romannumeral1}. INTRODUCTION}

Rare-earth based materials are fertile ground for studying a variety of interesting quantum phenomena, including quantum criticality, superconductivity (SC), and valence fluctuations \cite{Coleman_2007,RevModPhys.81.1551,Qimiao_2010,MichaelSmidman2023}.
In cerium, the unpaired 4$f$-electrons strongly hybridize with the conduction electrons \cite{Huang2019,Wu2021}, and it exhibits a range of unusual properties. Upon applying pressure at room temperature, cerium undergoes a structural transition from the face-centered cubic $\gamma$ phase to the isostructural $\alpha$ phase at about 0.8~GPa, which is accompanied by a sizable  (16\%) volume collapse \cite{Lawson1949}. However, the origin of the $\gamma$ $\rightarrow$ $\alpha$ transition is still under debate, where there are two main theoretical scenarios, namely a Mott transition \cite{Johansson1974} or Kondo volume collapse \cite{Allen1982}. Moreover, there exists a significant change in the Kondo hybridization between the localized 4$f$-electrons and conduction electrons across the transition \cite{2005s,2014,2015}. At around 4.5~GPa, the $\alpha$ phase transforms into either the $\alpha^{'}$ phase (orthorhombic) or the $\alpha^{''}$ phase (monoclinic), depending on how the sample is prepared. In particular, if the sample is mechanically deformed during preparation it will enter the $\alpha^{''}$ phase, otherwise it forms the  $\alpha^{'}$ phase \cite{McMahon1997,Dmitriev2004}. At pressures above 15~GPa, both the  $\alpha^{'}$ and  $\alpha^{''}$ phases transform to the body-centered tetragonal $\epsilon$ phase \cite{Vohra1999,Schiwek2002,Munro2020}. Recent electronic structure calculations of the high-pressure phases of cerium ($\alpha^{'}$ ,$\alpha^{''}$, and $\epsilon$) show that the electronic correlations in the $\alpha^{''}$ phase are stronger than in the other two phases, and the $\alpha^{''}$ phase has the largest electron effective masses ($m^*$ = 2.28$m_{\rm e}$ and 1.92$m_{\rm e}$ for the 4$f$-5/2 and 4$f$-7/2 components, respectively) \cite{Lu2018}.

SC in the $\alpha$ phase was reported with a $T_{\rm c}$ below 50~mK \cite{Probst1977,Probst19755}, while the $\alpha^{'}$ phase exhibits SC with a $T_{\rm c}$ of around 1.8~K at 5~GPa \cite{CESC1968}, which was attributed to a lattice-dynamical instability \cite{Loa2012}. However, transport studies of the high pressure phases of cerium are relatively rare \cite{1976R,Miyagawa2006,Sakigawa2007}, especially regarding the SC of the $\alpha^{''}$ phase. In particular, the interplay between SC and the multiple structural transitions remains unclear. This motivated us to extend the transport measurements of Ce up to pressures of 54~GPa. In this work we present a comprehensive study of the transport properties of cerium up to 54~GPa and construct the temperature-pressure phase diagram encompassing the various structural and superconducting phases. 

\section{\uppercase\expandafter{\romannumeral2}. EXPERIMENTAL METHODS}
The cerium samples used in this study were cut from an Alfa Aesar Ce ingot (99.9\%). Since the $\alpha^{''}$ phase is reported to appear by cold working \cite{McMahon1997}, we prepared our samples such that the $\alpha^{''}$ phase would be realized at around 4.5-15~GPa, thereby ensuring that the $\alpha$ $\rightarrow$ $\alpha^{''}$ and the $\alpha^{''}$ $\rightarrow$ $\epsilon$ transitions could be studied. Cerium samples were cut to pieces of approximate dimensions $120 \times 80 \times 20~\mu \mathrm{m}^3$ using a file in an oxygen-free glovebox, and were then loaded into a BeCu diamond anvil cell (DAC) with a 400-$\mu$m-diameter culet. A 100-$\mu$m-thick preindented rhenium gasket was covered with boron nitride for electrical insulation and a 200-$\mu$m-diameter hole was drilled as the sample chamber. Daphne oil 7373 was used as the pressure transmitting medium. The DAC was loaded together with several small ruby balls for pressure determination at room temperature using the ruby fluorescence method \cite{mao1986rubycalibration}. Electrical resistance measurements under pressure were performed in a Teslatron-PT system with an Oxford $^{3}$He refrigerator, across a temperature range of 0.3 to 300~K with a maximum applied magnetic field of 8~T. The ambient pressure resistivity measurements were performed using a Quantum Design Physical Property Measurement System (PPMS).

\begin{figure}[h]
	
	\includegraphics[width=1\columnwidth]{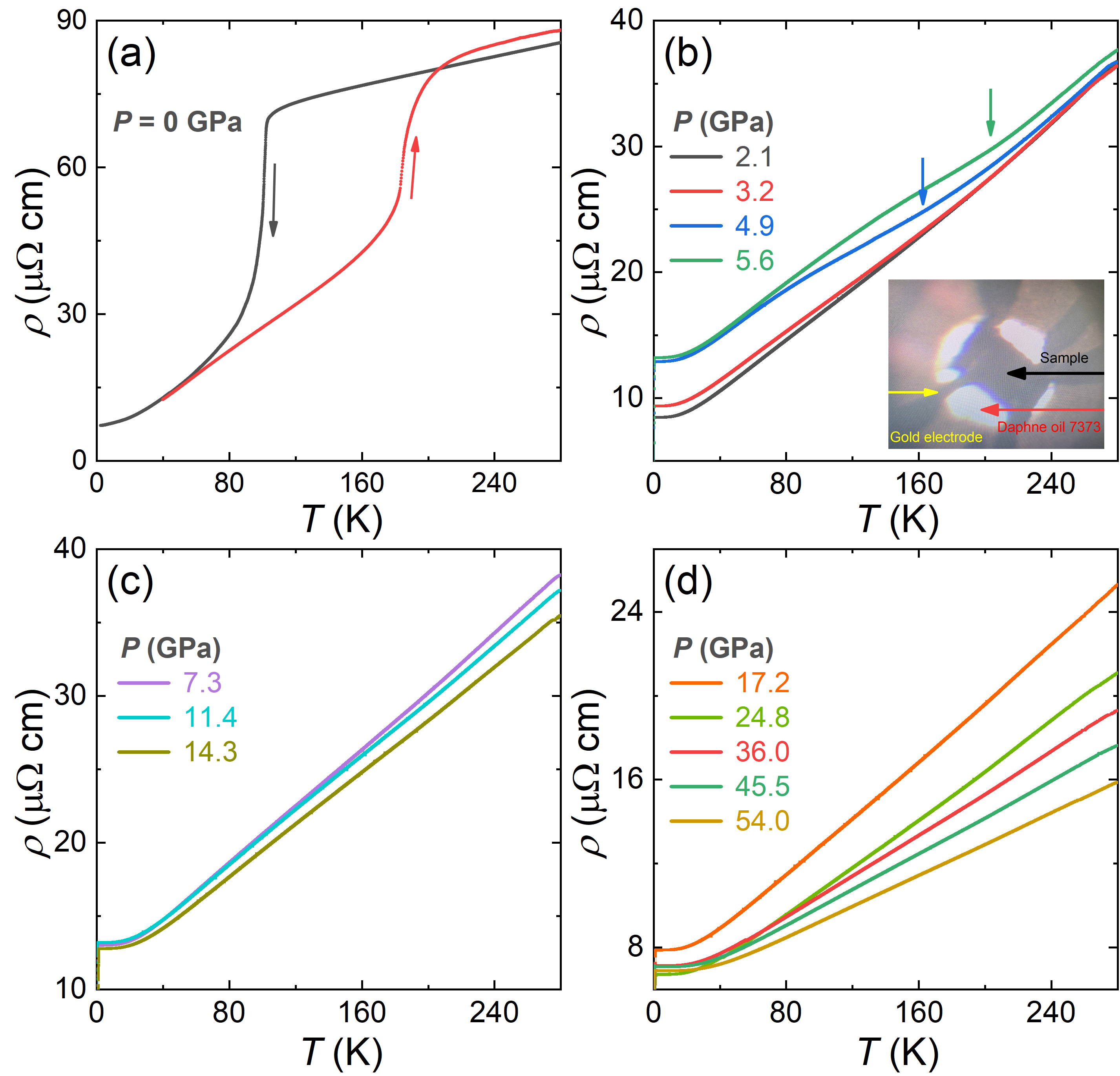}
	\caption{(Color online) (a) Resistivity of cerium sample at ambient pressure as a function of temperature $\rho$($T$) from 280~K down to 0.3~K. The black (red) arrow denotes the data taken upon cooling (warming). $\rho$($T$) is displayed under pressures of (b) 2.1 to 5.6~GPa, (c) 7.3 to 14.3~GPa, and (d) 17.2 to 54~GPa, where the blue and green arrows in (b) indicate the anomaly corresponding to the $\alpha$ $\rightarrow$ $\alpha^{''}$ transition. The inset in (b) shows the sample configuration in the DAC after pressure loading.}
	\label{fig1}
\end{figure}

\section{\uppercase\expandafter{\romannumeral3}. RESULTS}

Figure. \ref{fig1}(a) displays the temperature dependence of the resistivity $\rho$($T$) of cerium at ambient pressure from 280~K down to 0.3~K , which clearly exhibits an anomaly due to the isostructural  $\gamma$ $\rightarrow$ $\alpha$ transition. The large hysteresis loop between measurements performed upon cooling down and warming up indicates the first-order nature of this transition. 
\begin{figure}[b]
	\begin{center}
		\includegraphics[width=0.55\columnwidth]{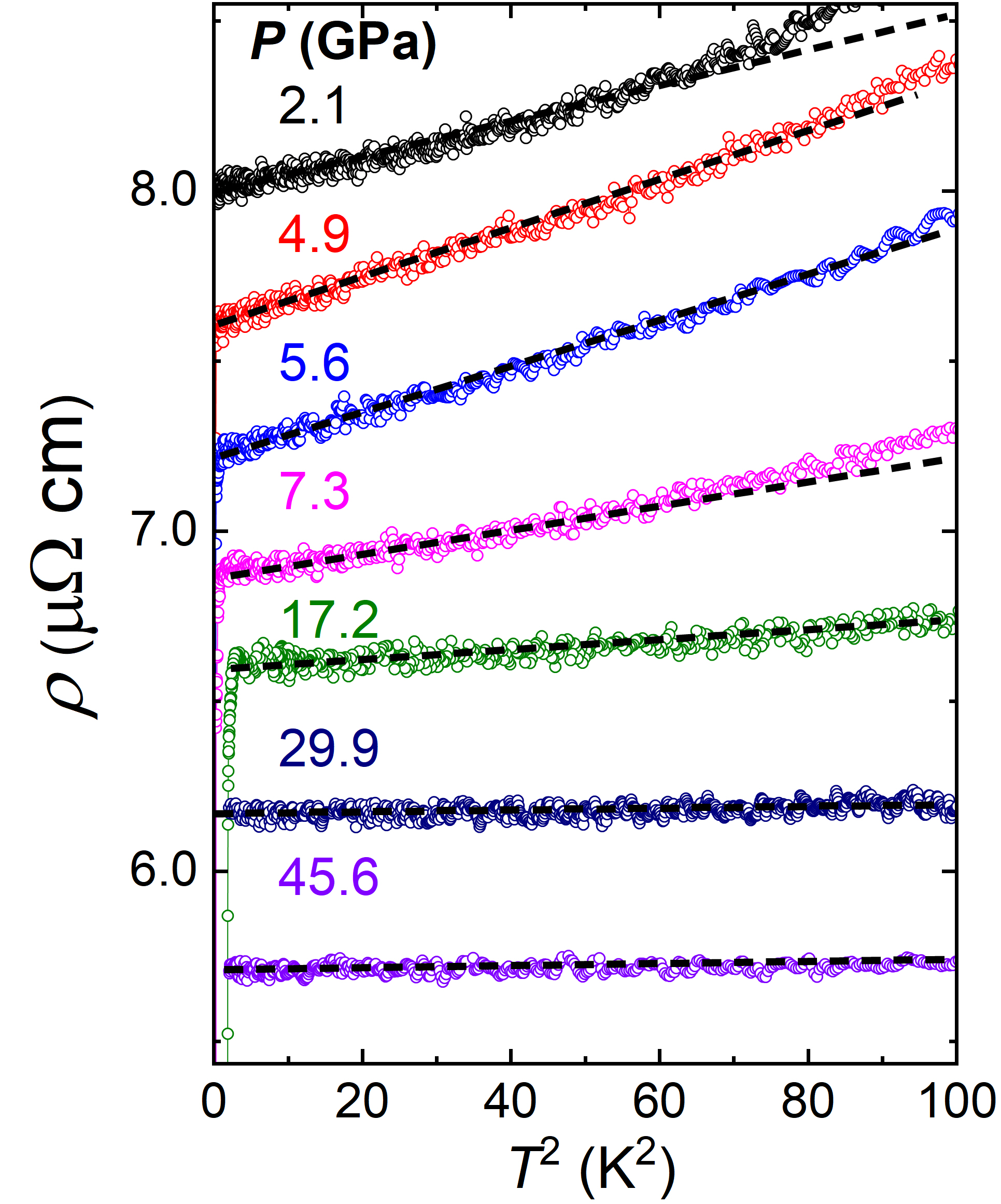}
	\end{center}
	\caption{(Color online) $\rho$($T$) versus $T^{\rm 2}$ at various pressures. The black dashed lines show fits to a $T^{\rm 2}$ dependence in the normal state, corresponding to Fermi liquid behavior. All the curves were shifted vertically for clarity.}
	\label{fig2}
\end{figure}

To investigate the interplay between the high-pressure structural transitions and SC, we performed resistivity measurements under pressure. $\rho$($T$) at selected pressures between 2.1~GPa and 54~GPa are displayed in Figs. \ref{fig1}(b)-\ref{fig1}(d). It can be seen that the $\gamma$ $\rightarrow$ $\alpha$ transition is absent at 2.1~GPa, which is consistent with previous transport results \cite{Miyagawa2006}. However, at a higher pressure of 4.9~GPa, an anomaly corresponding to the $\alpha$ $\rightarrow$ $\alpha^{''}$  transition \cite{McMahon1997} appears in the resistivity (blue arrow). This transition rapidly shifts to higher temperatures with increasing pressure, and exceeds 300~K when the pressure is above 5.6~GPa \cite{Schiwek2002}. Upon further increasing the pressure, as shown in Figs. \ref{fig1}(c) and \ref{fig1}(d), $\rho$($T$) exhibits a linear temperature dependence upon cooling from 280 K down to 60 K, reflecting the dominant role of electron–phonon scattering at high temperature \cite{ep,ep2}.

Meanwhile as shown in  Fig. \ref{fig2},  $\rho$($T$) at low temperatures ($T_{\rm c}$$<$ $T$ $<$ 8 K) can be fitted using $\rho$($T$) = $\rho$(0) + $A$$T^{\rm 2}$, where $\rho$(0) is the residual resistivity and $A$ is the coefficient related to Fermi-liquid behavior. The evolution of the $A$-coefficient with pressure exhibits three distinct behaviors as the pressure is changed. In the $\alpha$ phase (0.8 - 4.5~GPa), the $A$-coefficient gradually increases with pressure, reaching a maximum value of $5.3 \times$ $10^{-4} \mu \Omega~\mathrm{cm}~\mathrm{K}^{-2}$ at around 4.5 GPa. Conversely, in the $\alpha^{''}$ phase (4.5 - 15~GPa), there is a rapid decrease of the $A$-coefficient with increasing pressure, suggesting a possible pressure-induced suppression of the electron correlation strength. Meanwhile in the $\epsilon$ phase (above 15~GPa), the nearly temperature-independent resistivity implies fully  itinerant $f$ electrons. 

\begin{figure}[t]
	\begin{center}
		\includegraphics[width=0.95\columnwidth]{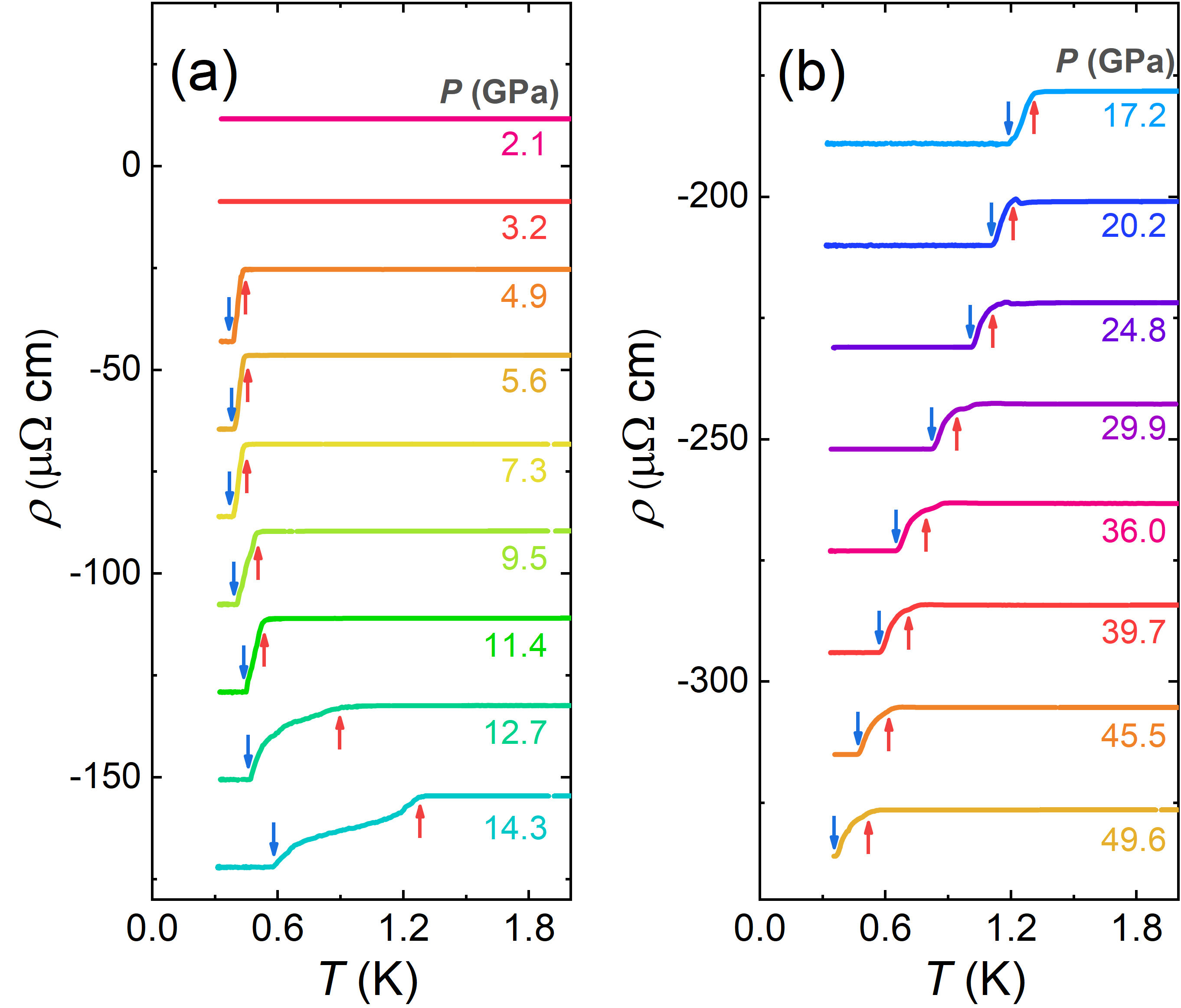}
	\end{center}
	\caption{(Color online) Low temperature $\rho$($T$) of cerium between (a) 2.1 and 14.3~GPa, and (b) 17.2 and 49.6~GPa. The red and blue arrows represent the $T_c^{\text {onset }}$ and $T_c^{\text {zero }}$, respectively. For clarity, all the curves in both panels are vertically  shifted relative to the 2.1 GPa curve.}
	\label{fig3}
\end{figure}

Figure. \ref{fig3} displays $\rho$($T$) under pressure at low temperatures in the vicinity of the superconducting transitions. As shown in Fig. \ref{fig3}(a),  there is no signature of SC down to 0.3~K in the $\alpha$ phase. Under a pressure of 4.9~GPa, the sample undergoes the $\alpha$ $\rightarrow$ $\alpha^{''}$ transition \cite{McMahon1997}, which is accompanied by the appearance of SC with a $T_{\rm c}$ of 0.4~K, and $T_{\rm c}$ slightly increases to 0.5~K at 11.4~GPa. Upon further increasing the pressure to 14.3~GPa, $T_{\rm c}$ exhibits a sudden increase once the $\alpha^{''}$ phase transforms into the $\epsilon$ phase. It is noted that the broad superconducting transitions observed at 12.7~GPa and 14.3~GPa are attributed to a slight pressure inhomogeneity, which may give  rise to a broad distribution of $T_{\rm c}$ while $T_{\rm c}$ is sharply enhanced across the structural phase transition.  In the $\epsilon$ phase, $T_{\rm c}$ monotonically decreases with increasing pressure up to at least 54~GPa. The pressure-induced changes of $T_{\rm c}$ are compatible with the structural phase  transitions determined by previous diffraction studies \cite{McMahon1997,Schiwek2002,Munro2020}. 

\begin{figure}[b]
	\begin{center}
		\includegraphics[width=1.\columnwidth]{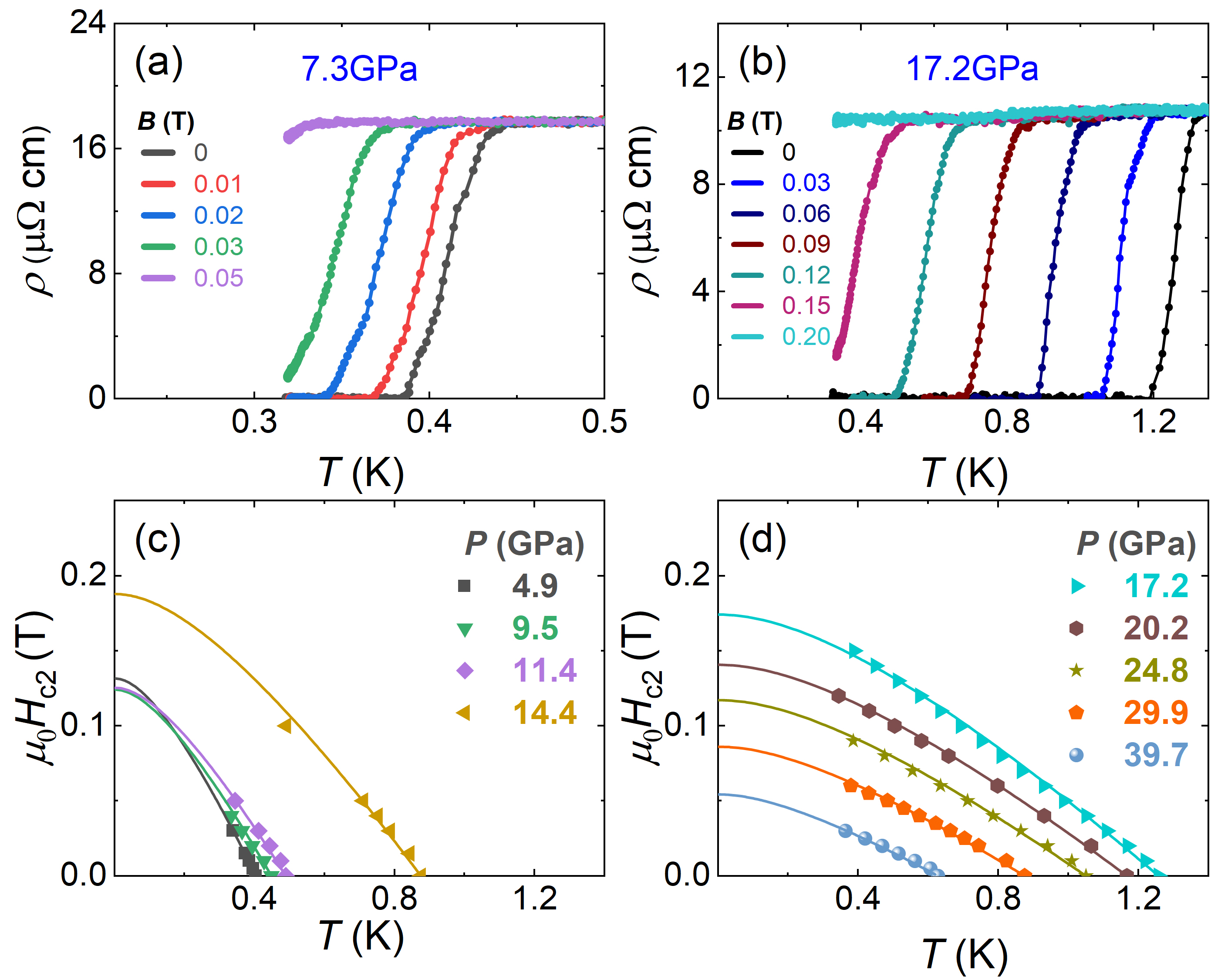}
	\end{center}
	\caption{(Color online) $\rho$($T$) under various applied magnetic fields at (a) 7.3~GPa and (b) 17.2~GPa. (c) , (d) The upper critical field $\mu_{\rm 0}H_{c2}$ versus temperature under various pressures. The solid lines show the results from fitting using the WHH model.}
	\label{fig4}
\end{figure}

To further probe the pressure-induced superconducting phases, we examined the effect of applied magnetic fields on $T_{\rm c}$. As an example, Figs. \ref{fig4}(a) and \ref{fig4}(b) display the superconducting transition under different magnetic fields at 7.3~GPa and 17.2~GPa respectively, where $T_{\rm c}$ is gradually suppressed by applied magnetic fields. Figures. \ref{fig4}(c) and (d) show the upper critical field $H_{c2}$($T$) (determined from the midpoint of the resistivity transitions) as a function of temperature at various pressures. $H_{c2}$($T$) are well fitted using the Werthamer-Helfand-Hohenberg (WHH) model \cite{WHH}, from which the zero temperature values $\mu_{\rm 0}H_{c2}$(0) are obtained, which show a significant pressure dependence. It can be seen that although $T_{\rm c}$ $\approx$ 0.9~K at 14.4~GPa is the same as the $T_{\rm c}$ at 29.9~GPa, SC is suppressed much more quickly under applied fields at 29.9~GPa, pointing to  a significant pressure tuning of the superconducting state not captured by
$T_{\rm c}$.

\begin{figure}[t]
	\includegraphics[angle=0,width=0.45\textwidth]{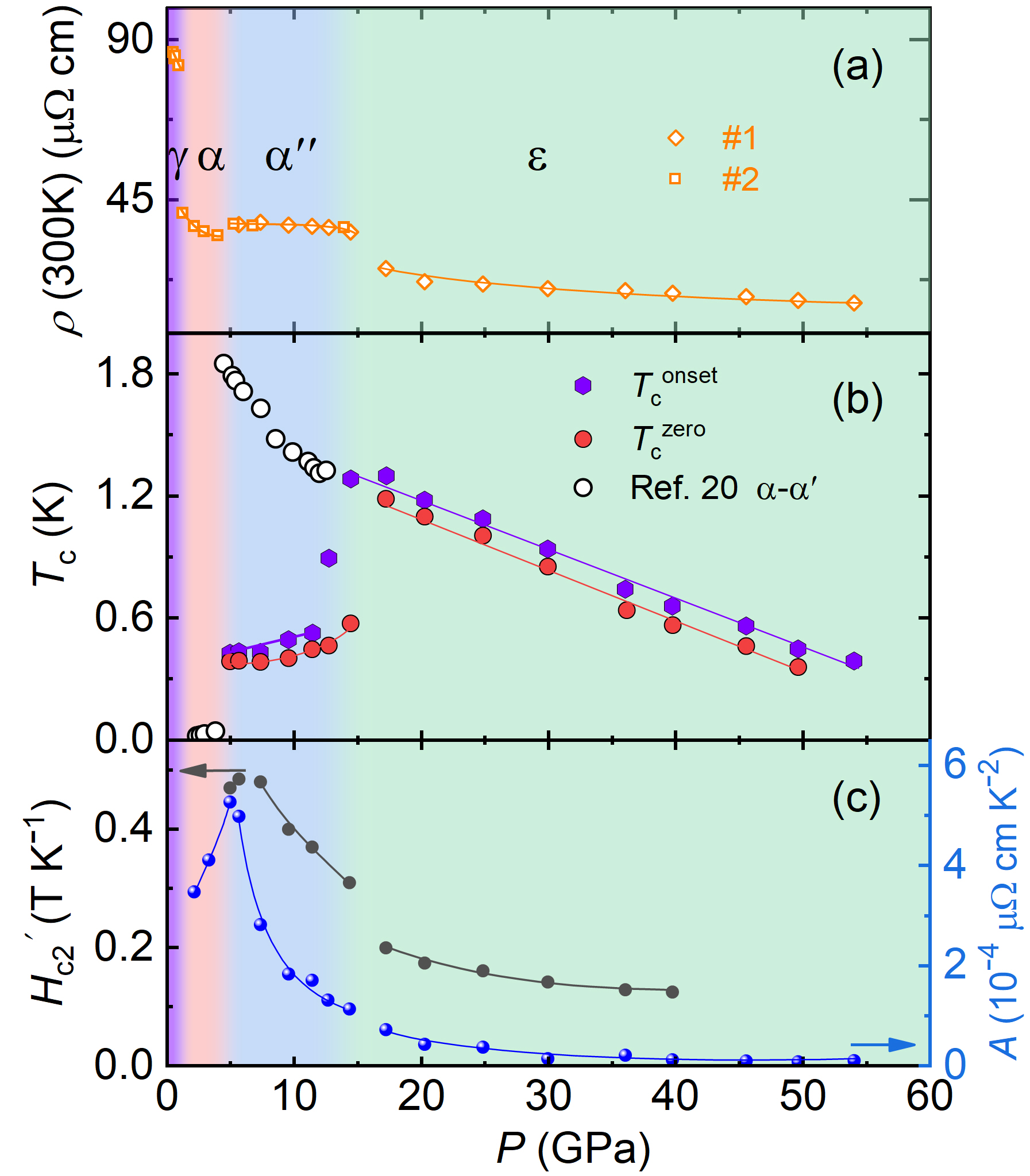}
	\vspace{-12pt} 
	\caption{(Color online)  Pressure dependence of (a) the room temperature resistivity $\rho$(300~K), (b) the superconducting transition temperature  $T_c^{\text {onset }}$ and $T_c^{\text {zero }}$, and (c) the initial slope of the upper critical field $H^{'}_{c2}$ ($H^{'}_{c2}$=$-$(d$H_{c2}$/d$T)_{T=T_{\rm c}}$) (black left axis) and the $A$-coefficient of the resistivity (blue right axis). The shaded regions correspond to different labelled structural phases. The white open circles in (b) are digitized from Ref. \cite{Probst19755}. }
	\vspace{-12pt}
	\label{fig5}
\end{figure}

In Fig. \ref{fig5}, we plot the temperature-pressure phase diagram up to 54 GPa showing the SC in the different structural phases. The four different structural phases, $\gamma$, $\alpha$, $\alpha^{''}$ and $\epsilon$, are evidenced from the pressure dependence of the room temperature resistivity  $\rho$(300~K) in Fig. \ref{fig5}(a), which exhibits three discontinuous changes. Firstly, $\rho$(300~K) jumps from 90~$\mu$$\Omega$~cm to 45~$\mu$$\Omega$~cm at around 0.8~GPa, and this large jump corresponds to the $\gamma$ $\rightarrow$ $\alpha$ transition. Secondly, there is a sudden increase at 4.5~GPa, which can be identified as the $\alpha$ $\rightarrow$ $\alpha^{''}$ transition. Finally, upon further compression to 15~GPa, another resistivity jump related to the $\alpha^{''}$ $\rightarrow$ $\epsilon$ transition is observed. All these features are highly consistent with previous diffraction studies \cite{McMahon1997,Schiwek2002,Munro2020}.

The evolution of SC with pressure is displayed in Fig. \ref{fig5}(b), where $T_{\rm c}$ decreases monotonically with increasing pressure in the $\alpha^{'}$ and $\epsilon$ phases, while it increases monotonically in the $\alpha^{''}$ phase. Moreover, the $\alpha^{''}$ $\rightarrow$ $\epsilon$ transition leads  to a significant enhancement of $T_{\rm c}$, and the maximum value of $T_{\rm c}$ (1.25~K) for the $\epsilon$ phase is achieved at 17.2~GPa, but the $T_{\rm c}$ does not change much when cerium undergoes the $\alpha^{'}$ $\rightarrow$ $\epsilon$ transition \cite{Probst19755}. The pressure dependence of the extracted initial slope of the upper critical field $H^{'}_{c2}$ ($H^{'}_{c2}$ = $-$(d$H_{c2}$/d$T)_{T=T_{\rm c}}$) and the $A$-coefficient are shown in Fig. \ref{fig5}(c), which both reach a maximum value near the critical pressure of the $\alpha$ $\rightarrow$ $\alpha^{''}$ transition and then undergo a rapid decrease in the $\alpha^{''}$ phase; the $A$-coefficient and the $H^{'}_{c2}$ decrease by a factor of 4.1 and 1.5 from 4.5~GPa to 15~GPa, respectively. 

\section{\uppercase\expandafter{\romannumeral4}. DISCUSSION AND SUMMARY}

In the $\alpha^{''}$ phase (4.5 - 15~GPa), theoretical calculations have identified electron-phonon coupling as the origin of SC and predicted a $T_{\rm c}$ of around 1~K at a pressure of 6.5~GPa \cite{Loa2012}. The calculated value of $T_{\rm c}$ is close to our experimental value of 0.4~K at 5~GPa. Moreover, it should be noted that there is a significant difference in the $T_{\rm c}$ between the $\alpha^{''}$ phase (0.5~K) measured here and $\alpha^{'}$ phase (1.8~K) at a pressure of 5~GPa \cite{Probst19755}, which can be attributed to differing strengths of the electron-phonon coupling \cite{Loa2012} and electronic correlations \cite{Lu2018}. 

Besides the value of $T_{\rm c}$, there is little information about the nature of the pressure-induced SC in cerium. As displayed in Fig. \ref{fig4}, the effect of applied magnetic fields suggests type II behavior with $\mu_{\rm 0}H_{c2}$(0) of 0.12~T at 4.9~GPa. Moreover, $T_{\rm c}$ shows an enhancement with increasing pressure in the $\alpha^{''}$ phase, which can be understood on very general grounds. The $\alpha^{''}$ phase appears to have stronger electronic correlations \cite{Lu2018}, and these correlations decrease with increasing pressure, leading to a slight increase of $T_{\rm c}$. Such a scenario is evidenced by the change of the $A$-coefficient and $H^{'}_{c2}$  in the $\alpha^{''}$ phase, from which we can estimate the change of effective carrier mass $m^*$ ($m^* \propto \sqrt{H_{\mathrm{c} 2}^{\prime}} / T_{\mathrm{c}}$, $m^* \propto \sqrt{A}$) \cite{knebel2008quantum,Park2008,BinShen2019}. Based on the change of $A$-coefficient from $5.3 \times$ $10^{-4}~\mu \Omega~\mathrm{cm}~\mathrm{K}^{-2}$ at  4.9 GPa to $1.3 \times$ $10^{-4}~\mu \Omega~\mathrm{cm}~\mathrm{K}^{-2}$ at 14.4 GPa, it can be estimated that $m^*$ decreases by almost a factor of two in this pressure range.

In contrast, in the $\epsilon$ phase (P $>$ 15~GPa), the 4$f$-electrons exhibit fully itinerant characteristics, as reflected in the near temperature-independent $\rho(T)$ from $T_{\rm c}$ to 10~K shown in Fig.~\ref{fig3}. This indicates a change from a state with sizeable electronic correlations, to one that is weakly correlated. Moreover, recent electronic structure calculations of the high pressure phases of cerium show that valence fluctuations in the $\epsilon$ phase are much stronger than in other phases \cite{Lu2018}. Such behavior has been found in Ce-based Kondo lattice systems such as {CeCu$_{2}$Si$_{2}$} \cite{Yuan2003,Yuan2006}, where there is a near temperature independent $\rho$($T$) in the high pressure regime. 

The sudden increase of $T_{\rm c}$ also coincides with the $\alpha^{''}$ $\rightarrow$ $\epsilon$ transition in a manner similar to the other elemental superconductors \cite{Hamlin2015,Zhang2022,chenSC}, suggesting a close correlation between pressure induced SC and structural phase transitions. Upon further increasing the pressure in the $\epsilon$ phase, the suppression of $T_{\rm c}$ with pressure occurs almost linearly up to 54~GPa (d$T_{\rm c}$/d$P$$ \approx -$32 mK/GPa). This is consistent with the expectations of conventional phonon-mediated SC, where $T_{\rm c}$ decreases at high pressures due to phonon stiffening \cite{MgB2}.

In conclusion, we systematically investigated the transport behaviors of cerium at high pressure and  constructed its superconducting phase diagram up to pressures of 54 GPa. Superconductivity in the $\alpha^{''}$ phase appears at around 4.9~GPa with a $T_{\rm c}$ of 0.4~K. The $A$-coefficient and $H^{'}_{c2}$ show a rapid decrease with increasing pressure in the $\alpha^{''}$ phase, which indicates a pressure-induced suppression of electronic correlations, that is accompanied by an increase of $T_{\rm c}$. Consequently, the evolution of superconductivity in cerium is closely related to the multiple pressure-induced structural transitions and the corresponding underlying electronic structures.

\section{ACKNOWLEDGMENTS}
We thank F. Steglich and C. Cao for interesting discussions. This work was supported by the National Key R\&D Program of China (Grant No. 2022YFA1402200), the Key R\&D Program of Zhejiang Province, China (Grant No. 2021C01002), the National Natural Science Foundation of China (Grants No. 12274364, No. 11974306, No. 12034017, and No. 12174332), and the Zhejiang Provincial Natural Science Foundation of China (Grant No. LR22A040002).

\bibliography{ref}

\begin{thebibliography}{38}%
\makeatletter
\providecommand \@ifxundefined [1]{%
 \@ifx{#1\undefined}
}%
\providecommand \@ifnum [1]{%
 \ifnum #1\expandafter \@firstoftwo
 \else \expandafter \@secondoftwo
 \fi
}%
\providecommand \@ifx [1]{%
 \ifx #1\expandafter \@firstoftwo
 \else \expandafter \@secondoftwo
 \fi
}%
\providecommand \natexlab [1]{#1}%
\providecommand \enquote  [1]{``#1''}%
\providecommand \bibnamefont  [1]{#1}%
\providecommand \bibfnamefont [1]{#1}%
\providecommand \citenamefont [1]{#1}%
\providecommand \href@noop [0]{\@secondoftwo}%
\providecommand \href [0]{\begingroup \@sanitize@url \@href}%
\providecommand \@href[1]{\@@startlink{#1}\@@href}%
\providecommand \@@href[1]{\endgroup#1\@@endlink}%
\providecommand \@sanitize@url [0]{\catcode `\\12\catcode `\$12\catcode
  `\&12\catcode `\#12\catcode `\^12\catcode `\_12\catcode `\%12\relax}%
\providecommand \@@startlink[1]{}%
\providecommand \@@endlink[0]{}%
\providecommand \url  [0]{\begingroup\@sanitize@url \@url }%
\providecommand \@url [1]{\endgroup\@href {#1}{\urlprefix }}%
\providecommand \urlprefix  [0]{URL }%
\providecommand \Eprint [0]{\href }%
\providecommand \doibase [0]{http://dx.doi.org/}%
\providecommand \selectlanguage [0]{\@gobble}%
\providecommand \bibinfo  [0]{\@secondoftwo}%
\providecommand \bibfield  [0]{\@secondoftwo}%
\providecommand \translation [1]{[#1]}%
\providecommand \BibitemOpen [0]{}%
\providecommand \bibitemStop [0]{}%
\providecommand \bibitemNoStop [0]{.\EOS\space}%
\providecommand \EOS [0]{\spacefactor3000\relax}%
\providecommand \BibitemShut  [1]{\csname bibitem#1\endcsname}%
\let\auto@bib@innerbib\@empty
\bibitem [{\citenamefont {Coleman}(2007)}]{Coleman_2007}%
  \BibitemOpen
  \bibfield  {author} {\bibinfo {author} {\bibfnamefont {Piers}\ \bibnamefont
  {Coleman}},\ }\href {\doibase 10.1002/9780470022184} {\emph {\bibinfo {title}
  {Handbook of Magnetism and Advanced Magnetic Materials}}}\ (\bibinfo
  {publisher} {Wiley},\ \bibinfo {year} {2007})\BibitemShut {NoStop}%
\bibitem [{\citenamefont {Pfleiderer}(2009)}]{RevModPhys.81.1551}%
  \BibitemOpen
  \bibfield  {author} {\bibinfo {author} {\bibfnamefont {Christian}\
  \bibnamefont {Pfleiderer}},\ }\bibfield  {title} {\enquote {\bibinfo {title}
  {Superconducting phases of $f$-electron compounds},}\ }\href {\doibase
  10.1103/RevModPhys.81.1551} {\bibfield  {journal} {\bibinfo  {journal} {Rev.
  Mod. Phys.}\ }\textbf {\bibinfo {volume} {81}},\ \bibinfo {pages}
  {1551--1624} (\bibinfo {year} {2009})}\BibitemShut {NoStop}%
\bibitem [{\citenamefont {Si}\ and\ \citenamefont
  {Steglich}(2010)}]{Qimiao_2010}%
  \BibitemOpen
  \bibfield  {author} {\bibinfo {author} {\bibfnamefont {Qimiao}\ \bibnamefont
  {Si}}\ and\ \bibinfo {author} {\bibfnamefont {Frank}\ \bibnamefont
  {Steglich}},\ }\bibfield  {title} {\enquote {\bibinfo {title} {Heavy fermions
  and quantum phase transitions},}\ }\href {\doibase 10.1126/science.1191195}
  {\bibfield  {journal} {\bibinfo  {journal} {Science}\ }\textbf {\bibinfo
  {volume} {329}},\ \bibinfo {pages} {1161--1166} (\bibinfo {year}
  {2010})}\BibitemShut {NoStop}%
\bibitem [{\citenamefont {Smidman}\ \emph {et~al.}(2023)\citenamefont
  {Smidman}, \citenamefont {Stockert}, \citenamefont {Nica}, \citenamefont
  {Liu}, \citenamefont {Yuan}, \citenamefont {Si},\ and\ \citenamefont
  {Steglich}}]{MichaelSmidman2023}%
  \BibitemOpen
  \bibfield  {author} {\bibinfo {author} {\bibfnamefont {Michael}\ \bibnamefont
  {Smidman}}, \bibinfo {author} {\bibfnamefont {Oliver}\ \bibnamefont
  {Stockert}}, \bibinfo {author} {\bibfnamefont {Emilian~M.}\ \bibnamefont
  {Nica}}, \bibinfo {author} {\bibfnamefont {Yang}\ \bibnamefont {Liu}},
  \bibinfo {author} {\bibfnamefont {Huiqiu}\ \bibnamefont {Yuan}}, \bibinfo
  {author} {\bibfnamefont {Qimiao}\ \bibnamefont {Si}}, \ and\ \bibinfo
  {author} {\bibfnamefont {Frank}\ \bibnamefont {Steglich}},\ }\href@noop {}
  {\enquote {\bibinfo {title} {Unconventional fully-gapped superconductivity in
  the heavy-fermion metal {CeCu$_2$Si$_2$}},}\ } (\bibinfo {year} {2023}),\
  \Eprint {http://arxiv.org/abs/2303.01901} {arXiv:2303.01901
  [cond-mat.supr-con]} \BibitemShut {NoStop}%
\bibitem [{\citenamefont {Huang}\ and\ \citenamefont {Lu}(2019)}]{Huang2019}%
  \BibitemOpen
  \bibfield  {author} {\bibinfo {author} {\bibfnamefont {Li}~\bibnamefont
  {Huang}}\ and\ \bibinfo {author} {\bibfnamefont {Haiyan}\ \bibnamefont
  {Lu}},\ }\bibfield  {title} {\enquote {\bibinfo {title} {Electronic structure
  of cerium: A comprehensive first-principles study},}\ }\href {\doibase
  10.1103/PhysRevB.99.045122} {\bibfield  {journal} {\bibinfo  {journal} {Phys.
  Rev. B}\ }\textbf {\bibinfo {volume} {99}},\ \bibinfo {pages} {045122}
  (\bibinfo {year} {2019})}\BibitemShut {NoStop}%
\bibitem [{\citenamefont {Wu}\ \emph {et~al.}(2021)\citenamefont {Wu},
  \citenamefont {Fang}, \citenamefont {Li}, \citenamefont {Xiao}, \citenamefont
  {Zheng}, \citenamefont {Yuan}, \citenamefont {Cao}, \citenamefont {feng
  Yang},\ and\ \citenamefont {Liu}}]{Wu2021}%
  \BibitemOpen
  \bibfield  {author} {\bibinfo {author} {\bibfnamefont {Yi}~\bibnamefont
  {Wu}}, \bibinfo {author} {\bibfnamefont {Yuan}\ \bibnamefont {Fang}},
  \bibinfo {author} {\bibfnamefont {Peng}\ \bibnamefont {Li}}, \bibinfo
  {author} {\bibfnamefont {Zhiguang}\ \bibnamefont {Xiao}}, \bibinfo {author}
  {\bibfnamefont {Hao}\ \bibnamefont {Zheng}}, \bibinfo {author} {\bibfnamefont
  {Huiqiu}\ \bibnamefont {Yuan}}, \bibinfo {author} {\bibfnamefont {Chao}\
  \bibnamefont {Cao}}, \bibinfo {author} {\bibfnamefont {Yi}~\bibnamefont {feng
  Yang}}, \ and\ \bibinfo {author} {\bibfnamefont {Yang}\ \bibnamefont {Liu}},\
  }\bibfield  {title} {\enquote {\bibinfo {title} {Bandwidth-control
  orbital-selective delocalization of 4$f$ electrons in epitaxial {Ce}
  films},}\ }\href {https://doi.org/10.1038/s41467-021-22710-2} {\bibfield
  {journal} {\bibinfo  {journal} {Nature Communications}\ }\textbf {\bibinfo
  {volume} {12}} (\bibinfo {year} {2021})}\BibitemShut {NoStop}%
\bibitem [{\citenamefont {Lawson}\ and\ \citenamefont
  {Tang}(1949)}]{Lawson1949}%
  \BibitemOpen
  \bibfield  {author} {\bibinfo {author} {\bibfnamefont {A.~W.}\ \bibnamefont
  {Lawson}}\ and\ \bibinfo {author} {\bibfnamefont {Ting-Yuan}\ \bibnamefont
  {Tang}},\ }\bibfield  {title} {\enquote {\bibinfo {title} {Concerning the
  high pressure allotropic modification of cerium},}\ }\href
  {https://doi.org/10.1103/physrev.76.301} {\bibfield  {journal} {\bibinfo
  {journal} {Physical Review}\ }\textbf {\bibinfo {volume} {76}},\ \bibinfo
  {pages} {301--302} (\bibinfo {year} {1949})}\BibitemShut {NoStop}%
\bibitem [{\citenamefont {Johansson}(1974)}]{Johansson1974}%
  \BibitemOpen
  \bibfield  {author} {\bibinfo {author} {\bibfnamefont {B\"{o}rje}\
  \bibnamefont {Johansson}},\ }\bibfield  {title} {\enquote {\bibinfo {title}
  {The $\alpha$-$\gamma$ transition in cerium is a mott transition},}\ }\href
  {\doibase 10.1080/14786439808206574} {\bibfield  {journal} {\bibinfo
  {journal} {Philosophical Magazine}\ }\textbf {\bibinfo {volume} {30}},\
  \bibinfo {pages} {469--482} (\bibinfo {year} {1974})}\BibitemShut {NoStop}%
\bibitem [{\citenamefont {Allen}\ and\ \citenamefont
  {Martin}(1982)}]{Allen1982}%
  \BibitemOpen
  \bibfield  {author} {\bibinfo {author} {\bibfnamefont {J.~W.}\ \bibnamefont
  {Allen}}\ and\ \bibinfo {author} {\bibfnamefont {Richard~M.}\ \bibnamefont
  {Martin}},\ }\bibfield  {title} {\enquote {\bibinfo {title} {Kondo volume
  collapse and the
  $\ensuremath{\gamma}\ensuremath{\rightarrow}\ensuremath{\alpha}$ transition
  in cerium},}\ }\href {\doibase 10.1103/PhysRevLett.49.1106} {\bibfield
  {journal} {\bibinfo  {journal} {Phys. Rev. Lett.}\ }\textbf {\bibinfo
  {volume} {49}},\ \bibinfo {pages} {1106--1110} (\bibinfo {year}
  {1982})}\BibitemShut {NoStop}%
\bibitem [{\citenamefont {Haule}\ \emph {et~al.}(2005)\citenamefont {Haule},
  \citenamefont {Oudovenko}, \citenamefont {Savrasov},\ and\ \citenamefont
  {Kotliar}}]{2005s}%
  \BibitemOpen
  \bibfield  {author} {\bibinfo {author} {\bibfnamefont {Kristjan}\
  \bibnamefont {Haule}}, \bibinfo {author} {\bibfnamefont {Viktor}\
  \bibnamefont {Oudovenko}}, \bibinfo {author} {\bibfnamefont {Sergej~Y.}\
  \bibnamefont {Savrasov}}, \ and\ \bibinfo {author} {\bibfnamefont {Gabriel}\
  \bibnamefont {Kotliar}},\ }\bibfield  {title} {\enquote {\bibinfo {title}
  {The $\ensuremath{\alpha}\ensuremath{\rightarrow}\ensuremath{\gamma}$
  transition in ce: A theoretical view from optical spectroscopy},}\ }\href
  {\doibase 10.1103/PhysRevLett.94.036401} {\bibfield  {journal} {\bibinfo
  {journal} {Phys. Rev. Lett.}\ }\textbf {\bibinfo {volume} {94}},\ \bibinfo
  {pages} {036401} (\bibinfo {year} {2005})}\BibitemShut {NoStop}%
\bibitem [{\citenamefont {Chakrabarti}\ \emph {et~al.}(2014)\citenamefont
  {Chakrabarti}, \citenamefont {Pezzoli}, \citenamefont {Sordi}, \citenamefont
  {Haule},\ and\ \citenamefont {Kotliar}}]{2014}%
  \BibitemOpen
  \bibfield  {author} {\bibinfo {author} {\bibfnamefont {B.}~\bibnamefont
  {Chakrabarti}}, \bibinfo {author} {\bibfnamefont {M.~E.}\ \bibnamefont
  {Pezzoli}}, \bibinfo {author} {\bibfnamefont {G.}~\bibnamefont {Sordi}},
  \bibinfo {author} {\bibfnamefont {K.}~\bibnamefont {Haule}}, \ and\ \bibinfo
  {author} {\bibfnamefont {G.}~\bibnamefont {Kotliar}},\ }\bibfield  {title}
  {\enquote {\bibinfo {title} {$\ensuremath{\alpha}$-$\ensuremath{\gamma}$
  transition in cerium: Magnetic form factor and dynamic magnetic
  susceptibility in dynamical mean-field theory},}\ }\href {\doibase
  10.1103/PhysRevB.89.125113} {\bibfield  {journal} {\bibinfo  {journal} {Phys.
  Rev. B}\ }\textbf {\bibinfo {volume} {89}},\ \bibinfo {pages} {125113}
  (\bibinfo {year} {2014})}\BibitemShut {NoStop}%
\bibitem [{\citenamefont {Haule}\ and\ \citenamefont {Birol}(2015)}]{2015}%
  \BibitemOpen
  \bibfield  {author} {\bibinfo {author} {\bibfnamefont {Kristjan}\
  \bibnamefont {Haule}}\ and\ \bibinfo {author} {\bibfnamefont {Turan}\
  \bibnamefont {Birol}},\ }\bibfield  {title} {\enquote {\bibinfo {title} {Free
  energy from stationary implementation of the $\mathrm{DFT}+\mathrm{DMFT}$
  functional},}\ }\href {\doibase 10.1103/PhysRevLett.115.256402} {\bibfield
  {journal} {\bibinfo  {journal} {Phys. Rev. Lett.}\ }\textbf {\bibinfo
  {volume} {115}},\ \bibinfo {pages} {256402} (\bibinfo {year}
  {2015})}\BibitemShut {NoStop}%
\bibitem [{\citenamefont {McMahon}\ and\ \citenamefont
  {Nelmes}(1997)}]{McMahon1997}%
  \BibitemOpen
  \bibfield  {author} {\bibinfo {author} {\bibfnamefont {M.~I.}\ \bibnamefont
  {McMahon}}\ and\ \bibinfo {author} {\bibfnamefont {R.~J.}\ \bibnamefont
  {Nelmes}},\ }\bibfield  {title} {\enquote {\bibinfo {title} {Different
  results for the equilibrium phases of cerium above {5~GPa}},}\ }\href
  {https://doi.org/10.1103/physrevlett.78.3884} {\bibfield  {journal} {\bibinfo
   {journal} {Phys. Rev. Lett.}\ }\textbf {\bibinfo {volume} {78}},\ \bibinfo
  {pages} {3884--3887} (\bibinfo {year} {1997})}\BibitemShut {NoStop}%
\bibitem [{\citenamefont {Dmitriev}\ \emph {et~al.}(2004)\citenamefont
  {Dmitriev}, \citenamefont {Kuznetsov}, \citenamefont {Bandilet},
  \citenamefont {Bouvier}, \citenamefont {Dubrovinsky}, \citenamefont
  {Machon},\ and\ \citenamefont {Weber}}]{Dmitriev2004}%
  \BibitemOpen
  \bibfield  {author} {\bibinfo {author} {\bibfnamefont {V.~P.}\ \bibnamefont
  {Dmitriev}}, \bibinfo {author} {\bibfnamefont {A.~Yu.}\ \bibnamefont
  {Kuznetsov}}, \bibinfo {author} {\bibfnamefont {O.}~\bibnamefont {Bandilet}},
  \bibinfo {author} {\bibfnamefont {P.}~\bibnamefont {Bouvier}}, \bibinfo
  {author} {\bibfnamefont {L.}~\bibnamefont {Dubrovinsky}}, \bibinfo {author}
  {\bibfnamefont {D.}~\bibnamefont {Machon}}, \ and\ \bibinfo {author}
  {\bibfnamefont {H.-P.}\ \bibnamefont {Weber}},\ }\bibfield  {title} {\enquote
  {\bibinfo {title} {Stability of the high-pressure monoclinic phases in
  $\mathrm{Ce}$ and $\mathrm{Pr}$ metals: Comparative diffraction study and
  phenomenological theory},}\ }\href {\doibase 10.1103/PhysRevB.70.014104}
  {\bibfield  {journal} {\bibinfo  {journal} {Phys. Rev. B}\ }\textbf {\bibinfo
  {volume} {70}},\ \bibinfo {pages} {014104} (\bibinfo {year}
  {2004})}\BibitemShut {NoStop}%
\bibitem [{\citenamefont {Vohra}\ \emph {et~al.}(1999)\citenamefont {Vohra},
  \citenamefont {Beaver}, \citenamefont {Akella}, \citenamefont {Ruddle},\ and\
  \citenamefont {Weir}}]{Vohra1999}%
  \BibitemOpen
  \bibfield  {author} {\bibinfo {author} {\bibfnamefont {Yogesh~K.}\
  \bibnamefont {Vohra}}, \bibinfo {author} {\bibfnamefont {Steven~L.}\
  \bibnamefont {Beaver}}, \bibinfo {author} {\bibfnamefont {Jagannadham}\
  \bibnamefont {Akella}}, \bibinfo {author} {\bibfnamefont {Chantel~A.}\
  \bibnamefont {Ruddle}}, \ and\ \bibinfo {author} {\bibfnamefont {Samuel~T.}\
  \bibnamefont {Weir}},\ }\bibfield  {title} {\enquote {\bibinfo {title}
  {Ultrapressure equation of state of cerium metal to {208~GPa}},}\ }\href
  {https://doi.org/10.1063/1.369566} {\bibfield  {journal} {\bibinfo  {journal}
  {Journal of Applied Physics}\ }\textbf {\bibinfo {volume} {85}},\ \bibinfo
  {pages} {2451--2453} (\bibinfo {year} {1999})}\BibitemShut {NoStop}%
\bibitem [{\citenamefont {Schiwek}\ \emph {et~al.}(2002)\citenamefont
  {Schiwek}, \citenamefont {Porsch},\ and\ \citenamefont
  {Holzapfel}}]{Schiwek2002}%
  \BibitemOpen
  \bibfield  {author} {\bibinfo {author} {\bibfnamefont {A.}~\bibnamefont
  {Schiwek}}, \bibinfo {author} {\bibfnamefont {F.}~\bibnamefont {Porsch}}, \
  and\ \bibinfo {author} {\bibfnamefont {W.~B.}\ \bibnamefont {Holzapfel}},\
  }\bibfield  {title} {\enquote {\bibinfo {title} {High temperature-high
  pressure structural studies of cerium},}\ }\href {\doibase
  10.1080/08957950212799} {\bibfield  {journal} {\bibinfo  {journal} {High
  Pressure Research}\ }\textbf {\bibinfo {volume} {22}},\ \bibinfo {pages}
  {407--410} (\bibinfo {year} {2002})}\BibitemShut {NoStop}%
\bibitem [{\citenamefont {Munro}\ \emph {et~al.}(2020)\citenamefont {Munro},
  \citenamefont {Daisenberger}, \citenamefont {MacLeod}, \citenamefont
  {McGuire}, \citenamefont {Loa}, \citenamefont {Popescu}, \citenamefont
  {Botella}, \citenamefont {Errandonea},\ and\ \citenamefont
  {McMahon}}]{Munro2020}%
  \BibitemOpen
  \bibfield  {author} {\bibinfo {author} {\bibfnamefont {K~A}\ \bibnamefont
  {Munro}}, \bibinfo {author} {\bibfnamefont {D}~\bibnamefont {Daisenberger}},
  \bibinfo {author} {\bibfnamefont {S~G}\ \bibnamefont {MacLeod}}, \bibinfo
  {author} {\bibfnamefont {S}~\bibnamefont {McGuire}}, \bibinfo {author}
  {\bibfnamefont {I}~\bibnamefont {Loa}}, \bibinfo {author} {\bibfnamefont
  {C}~\bibnamefont {Popescu}}, \bibinfo {author} {\bibfnamefont
  {P}~\bibnamefont {Botella}}, \bibinfo {author} {\bibfnamefont
  {D}~\bibnamefont {Errandonea}}, \ and\ \bibinfo {author} {\bibfnamefont
  {M~I}\ \bibnamefont {McMahon}},\ }\bibfield  {title} {\enquote {\bibinfo
  {title} {The high-pressure, high-temperature phase diagram of cerium},}\
  }\href {\doibase 10.1088/1361-648X/ab7f02} {\bibfield  {journal} {\bibinfo
  {journal} {Journal of Physics: Condensed Matter}\ }\textbf {\bibinfo {volume}
  {32}},\ \bibinfo {pages} {335401} (\bibinfo {year} {2020})}\BibitemShut
  {NoStop}%
\bibitem [{\citenamefont {Lu}\ and\ \citenamefont {Huang}(2018)}]{Lu2018}%
  \BibitemOpen
  \bibfield  {author} {\bibinfo {author} {\bibfnamefont {Haiyan}\ \bibnamefont
  {Lu}}\ and\ \bibinfo {author} {\bibfnamefont {Li}~\bibnamefont {Huang}},\
  }\bibfield  {title} {\enquote {\bibinfo {title} {Electronic correlations in
  cerium's high-pressure phases},}\ }\href {\doibase 10.1088/1361-648x/aadc7c}
  {\bibfield  {journal} {\bibinfo  {journal} {Journal of Physics: Condensed
  Matter}\ }\textbf {\bibinfo {volume} {30}},\ \bibinfo {pages} {395601}
  (\bibinfo {year} {2018})}\BibitemShut {NoStop}%
\bibitem [{\citenamefont {Probst}\ and\ \citenamefont
  {Wittig}(1977)}]{Probst1977}%
  \BibitemOpen
  \bibfield  {author} {\bibinfo {author} {\bibfnamefont {C.}~\bibnamefont
  {Probst}}\ and\ \bibinfo {author} {\bibfnamefont {J.}~\bibnamefont
  {Wittig}},\ }\bibfield  {title} {\enquote {\bibinfo {title}
  {Superconductivity in cerium at high pressure: Indications for phonon
  softening},}\ }\href {\doibase 10.1080/00150197708237175} {\bibfield
  {journal} {\bibinfo  {journal} {Ferroelectrics}\ }\textbf {\bibinfo {volume}
  {16}},\ \bibinfo {pages} {267--267} (\bibinfo {year} {1977})}\BibitemShut
  {NoStop}%
\bibitem [{\citenamefont {Probst}\ and\ \citenamefont
  {Wittig}(1975)}]{Probst19755}%
  \BibitemOpen
  \bibfield  {author} {\bibinfo {author} {\bibfnamefont {C.}~\bibnamefont
  {Probst}}\ and\ \bibinfo {author} {\bibfnamefont {J.}~\bibnamefont
  {Wittig}},\ }\bibfield  {title} {\enquote {\bibinfo {title}
  {Superconductivity of alpha cerium},}\ }\href
  {https://www.scopus.com/inward/record.uri?eid=2-s2.0-84973249091&partnerID=40&md5=c9895905d5b867a45a7a48ce44886400}
  {\bibfield  {journal} {\bibinfo  {journal} {Proceedings of the 14th
  International Conference on Low Temperature Physics (LT14)}\ ,\ \bibinfo
  {pages} {5}} (\bibinfo {year} {1975})}\BibitemShut {NoStop}%
\bibitem [{\citenamefont {Wittig}(1968)}]{CESC1968}%
  \BibitemOpen
  \bibfield  {author} {\bibinfo {author} {\bibfnamefont {J\"org}\ \bibnamefont
  {Wittig}},\ }\bibfield  {title} {\enquote {\bibinfo {title}
  {Superconductivity of cerium under pressure},}\ }\href {\doibase
  10.1103/PhysRevLett.21.1250} {\bibfield  {journal} {\bibinfo  {journal}
  {Phys. Rev. Lett.}\ }\textbf {\bibinfo {volume} {21}},\ \bibinfo {pages}
  {1250--1252} (\bibinfo {year} {1968})}\BibitemShut {NoStop}%
\bibitem [{\citenamefont {Loa}\ \emph {et~al.}(2012)\citenamefont {Loa},
  \citenamefont {Isaev}, \citenamefont {McMahon}, \citenamefont {Kim},
  \citenamefont {Johansson}, \citenamefont {Bosak},\ and\ \citenamefont
  {Krisch}}]{Loa2012}%
  \BibitemOpen
  \bibfield  {author} {\bibinfo {author} {\bibfnamefont {I.}~\bibnamefont
  {Loa}}, \bibinfo {author} {\bibfnamefont {E.~I.}\ \bibnamefont {Isaev}},
  \bibinfo {author} {\bibfnamefont {M.~I.}\ \bibnamefont {McMahon}}, \bibinfo
  {author} {\bibfnamefont {D.~Y.}\ \bibnamefont {Kim}}, \bibinfo {author}
  {\bibfnamefont {B.}~\bibnamefont {Johansson}}, \bibinfo {author}
  {\bibfnamefont {A.}~\bibnamefont {Bosak}}, \ and\ \bibinfo {author}
  {\bibfnamefont {M.}~\bibnamefont {Krisch}},\ }\bibfield  {title} {\enquote
  {\bibinfo {title} {Lattice dynamics and superconductivity in cerium at high
  pressure},}\ }\href {\doibase 10.1103/PhysRevLett.108.045502} {\bibfield
  {journal} {\bibinfo  {journal} {Phys. Rev. Lett.}\ }\textbf {\bibinfo
  {volume} {108}},\ \bibinfo {pages} {045502} (\bibinfo {year}
  {2012})}\BibitemShut {NoStop}%
\bibitem [{\citenamefont {Leger}(1976)}]{1976R}%
  \BibitemOpen
  \bibfield  {author} {\bibinfo {author} {\bibfnamefont {JM}~\bibnamefont
  {Leger}},\ }\bibfield  {title} {\enquote {\bibinfo {title} {Electrical
  resistivity of $\alpha$ cerium under high pressure},}\ }\href@noop {}
  {\bibfield  {journal} {\bibinfo  {journal} {Physics Letters A}\ }\textbf
  {\bibinfo {volume} {57}},\ \bibinfo {pages} {191--192} (\bibinfo {year}
  {1976})}\BibitemShut {NoStop}%
\bibitem [{\citenamefont {Miyagawa}\ \emph {et~al.}(2006)\citenamefont
  {Miyagawa}, \citenamefont {Oomi}, \citenamefont {Ohashi}, \citenamefont
  {Maezawa},\ and\ \citenamefont {Kagayama}}]{Miyagawa2006}%
  \BibitemOpen
  \bibfield  {author} {\bibinfo {author} {\bibfnamefont {H.}~\bibnamefont
  {Miyagawa}}, \bibinfo {author} {\bibfnamefont {G.}~\bibnamefont {Oomi}},
  \bibinfo {author} {\bibfnamefont {M.}~\bibnamefont {Ohashi}}, \bibinfo
  {author} {\bibfnamefont {K.}~\bibnamefont {Maezawa}}, \ and\ \bibinfo
  {author} {\bibfnamefont {T.}~\bibnamefont {Kagayama}},\ }\bibfield  {title}
  {\enquote {\bibinfo {title} {Pressure-enhanced magnetoresistance of -{Ce}
  single crystal},}\ }\href {\doibase 10.1016/j.jallcom.2005.04.098} {\bibfield
   {journal} {\bibinfo  {journal} {Journal of Alloys and Compounds}\ }\textbf
  {\bibinfo {volume} {408-412}},\ \bibinfo {pages} {230--233} (\bibinfo {year}
  {2006})}\BibitemShut {NoStop}%
\bibitem [{\citenamefont {Sakigawa}\ \emph {et~al.}(2007)\citenamefont
  {Sakigawa}, \citenamefont {Ohashi},\ and\ \citenamefont
  {Oomi}}]{Sakigawa2007}%
  \BibitemOpen
  \bibfield  {author} {\bibinfo {author} {\bibfnamefont {Yukio}\ \bibnamefont
  {Sakigawa}}, \bibinfo {author} {\bibfnamefont {Masashi}\ \bibnamefont
  {Ohashi}}, \ and\ \bibinfo {author} {\bibfnamefont {Gendo}\ \bibnamefont
  {Oomi}},\ }\bibfield  {title} {\enquote {\bibinfo {title} {Effect of pressure
  on electrical resistivity and magnetoresistance of $\beta$-{Ce}},}\ }\href
  {\doibase 10.1143/jpsjs.76sa.66} {\bibfield  {journal} {\bibinfo  {journal}
  {Journal of the Physical Society of Japan}\ }\textbf {\bibinfo {volume}
  {76}},\ \bibinfo {pages} {66--67} (\bibinfo {year} {2007})}\BibitemShut
  {NoStop}%
\bibitem [{\citenamefont {Mao}\ \emph {et~al.}(1986)\citenamefont {Mao},
  \citenamefont {Xu},\ and\ \citenamefont {Bell}}]{mao1986rubycalibration}%
  \BibitemOpen
  \bibfield  {author} {\bibinfo {author} {\bibfnamefont {H.~K.}\ \bibnamefont
  {Mao}}, \bibinfo {author} {\bibfnamefont {J.}~\bibnamefont {Xu}}, \ and\
  \bibinfo {author} {\bibfnamefont {P.~M.}\ \bibnamefont {Bell}},\ }\bibfield
  {title} {\enquote {\bibinfo {title} {Calibration of the ruby pressure gauge
  to 800 kbar under quasi-hydrostatic conditions},}\ }\href {\doibase
  10.1029/jb091ib05p04673} {\bibfield  {journal} {\bibinfo  {journal} {Journal
  of Geophysical Research}\ }\textbf {\bibinfo {volume} {91}},\ \bibinfo
  {pages} {4673} (\bibinfo {year} {1986})}\BibitemShut {NoStop}%
\bibitem [{\citenamefont {Plombon}\ \emph {et~al.}(2006)\citenamefont
  {Plombon}, \citenamefont {Andideh}, \citenamefont {Dubin},\ and\
  \citenamefont {Maiz}}]{ep}%
  \BibitemOpen
  \bibfield  {author} {\bibinfo {author} {\bibfnamefont {J.~J.}\ \bibnamefont
  {Plombon}}, \bibinfo {author} {\bibfnamefont {Ebrahim}\ \bibnamefont
  {Andideh}}, \bibinfo {author} {\bibfnamefont {Valery~M.}\ \bibnamefont
  {Dubin}}, \ and\ \bibinfo {author} {\bibfnamefont {Jose}\ \bibnamefont
  {Maiz}},\ }\bibfield  {title} {\enquote {\bibinfo {title} {Influence of
  phonon, geometry, impurity, and grain size on copper line resistivity},}\
  }\href {\doibase 10.1063/1.2355435} {\bibfield  {journal} {\bibinfo
  {journal} {Applied Physics Letters}\ }\textbf {\bibinfo {volume} {89}},\
  \bibinfo {pages} {113124} (\bibinfo {year} {2006})}\BibitemShut {NoStop}%
\bibitem [{\citenamefont {Polshyn}\ \emph {et~al.}(2019)\citenamefont
  {Polshyn}, \citenamefont {Yankowitz}, \citenamefont {Chen}, \citenamefont
  {Zhang}, \citenamefont {Watanabe}, \citenamefont {Taniguchi}, \citenamefont
  {Dean},\ and\ \citenamefont {Young}}]{ep2}%
  \BibitemOpen
  \bibfield  {author} {\bibinfo {author} {\bibfnamefont {Hryhoriy}\
  \bibnamefont {Polshyn}}, \bibinfo {author} {\bibfnamefont {Matthew}\
  \bibnamefont {Yankowitz}}, \bibinfo {author} {\bibfnamefont {Shaowen}\
  \bibnamefont {Chen}}, \bibinfo {author} {\bibfnamefont {Yuxuan}\ \bibnamefont
  {Zhang}}, \bibinfo {author} {\bibfnamefont {K.}~\bibnamefont {Watanabe}},
  \bibinfo {author} {\bibfnamefont {T.}~\bibnamefont {Taniguchi}}, \bibinfo
  {author} {\bibfnamefont {Cory~R.}\ \bibnamefont {Dean}}, \ and\ \bibinfo
  {author} {\bibfnamefont {Andrea~F.}\ \bibnamefont {Young}},\ }\bibfield
  {title} {\enquote {\bibinfo {title} {Large linear-in-temperature resistivity
  in twisted bilayer graphene},}\ }\href {\doibase 10.1038/s41567-019-0596-3}
  {\bibfield  {journal} {\bibinfo  {journal} {Nature Physics}\ }\textbf
  {\bibinfo {volume} {15}},\ \bibinfo {pages} {1011--1016} (\bibinfo {year}
  {2019})}\BibitemShut {NoStop}%
\bibitem [{\citenamefont {Werthamer}\ \emph {et~al.}(1966)\citenamefont
  {Werthamer}, \citenamefont {Helfand},\ and\ \citenamefont {Hohenberg}}]{WHH}%
  \BibitemOpen
  \bibfield  {author} {\bibinfo {author} {\bibfnamefont {N.~R.}\ \bibnamefont
  {Werthamer}}, \bibinfo {author} {\bibfnamefont {E.}~\bibnamefont {Helfand}},
  \ and\ \bibinfo {author} {\bibfnamefont {P.~C.}\ \bibnamefont {Hohenberg}},\
  }\bibfield  {title} {\enquote {\bibinfo {title} {Temperature and purity
  dependence of the superconducting critical field, ${H}_{c2}$. iii. electron
  spin and spin-orbit effects},}\ }\href {\doibase 10.1103/PhysRev.147.295}
  {\bibfield  {journal} {\bibinfo  {journal} {Phys. Rev.}\ }\textbf {\bibinfo
  {volume} {147}},\ \bibinfo {pages} {295--302} (\bibinfo {year}
  {1966})}\BibitemShut {NoStop}%
\bibitem [{\citenamefont {Knebel}\ \emph {et~al.}(2008)\citenamefont {Knebel},
  \citenamefont {Aoki}, \citenamefont {Brison},\ and\ \citenamefont
  {Flouquet}}]{knebel2008quantum}%
  \BibitemOpen
  \bibfield  {author} {\bibinfo {author} {\bibfnamefont {Georg}\ \bibnamefont
  {Knebel}}, \bibinfo {author} {\bibfnamefont {Dai}\ \bibnamefont {Aoki}},
  \bibinfo {author} {\bibfnamefont {Jean-Pascal}\ \bibnamefont {Brison}}, \
  and\ \bibinfo {author} {\bibfnamefont {Jacques}\ \bibnamefont {Flouquet}},\
  }\bibfield  {title} {\enquote {\bibinfo {title} {The quantum critical point
  in {CeRhIn$_{5}$}: A resistivity study},}\ }\href {\doibase
  10.1143/jpsj.77.114704} {\bibfield  {journal} {\bibinfo  {journal} {Journal
  of the Physical Society of Japan}\ }\textbf {\bibinfo {volume} {77}},\
  \bibinfo {pages} {114704} (\bibinfo {year} {2008})}\BibitemShut {NoStop}%
\bibitem [{\citenamefont {Park}\ \emph {et~al.}(2008)\citenamefont {Park},
  \citenamefont {Graf}, \citenamefont {Boulaevskii}, \citenamefont {Sarrao},\
  and\ \citenamefont {Thompson}}]{Park2008}%
  \BibitemOpen
  \bibfield  {author} {\bibinfo {author} {\bibfnamefont {T.}~\bibnamefont
  {Park}}, \bibinfo {author} {\bibfnamefont {M.~J.}\ \bibnamefont {Graf}},
  \bibinfo {author} {\bibfnamefont {L.}~\bibnamefont {Boulaevskii}}, \bibinfo
  {author} {\bibfnamefont {J.~L.}\ \bibnamefont {Sarrao}}, \ and\ \bibinfo
  {author} {\bibfnamefont {J.~D.}\ \bibnamefont {Thompson}},\ }\bibfield
  {title} {\enquote {\bibinfo {title} {Electronic duality in strongly
  correlated matter},}\ }\href {\doibase 10.1073/pnas.0801873105} {\bibfield
  {journal} {\bibinfo  {journal} {Proceedings of the National Academy of
  Sciences}\ }\textbf {\bibinfo {volume} {105}},\ \bibinfo {pages} {6825--6828}
  (\bibinfo {year} {2008})}\BibitemShut {NoStop}%
\bibitem [{\citenamefont {Shen}\ \emph {et~al.}(2020)\citenamefont {Shen},
  \citenamefont {Zhang}, \citenamefont {Komijani}, \citenamefont {Nicklas},
  \citenamefont {Borth}, \citenamefont {Wang}, \citenamefont {Chen},
  \citenamefont {Nie}, \citenamefont {Li}, \citenamefont {Lu}, \citenamefont
  {Lee}, \citenamefont {Smidman}, \citenamefont {Steglich}, \citenamefont
  {Coleman},\ and\ \citenamefont {Yuan}}]{BinShen2019}%
  \BibitemOpen
  \bibfield  {author} {\bibinfo {author} {\bibfnamefont {Bin}\ \bibnamefont
  {Shen}}, \bibinfo {author} {\bibfnamefont {Yongjun}\ \bibnamefont {Zhang}},
  \bibinfo {author} {\bibfnamefont {Yashar}\ \bibnamefont {Komijani}}, \bibinfo
  {author} {\bibfnamefont {Michael}\ \bibnamefont {Nicklas}}, \bibinfo {author}
  {\bibfnamefont {Robert}\ \bibnamefont {Borth}}, \bibinfo {author}
  {\bibfnamefont {An}~\bibnamefont {Wang}}, \bibinfo {author} {\bibfnamefont
  {Ye}~\bibnamefont {Chen}}, \bibinfo {author} {\bibfnamefont {Zhiyong}\
  \bibnamefont {Nie}}, \bibinfo {author} {\bibfnamefont {Rui}\ \bibnamefont
  {Li}}, \bibinfo {author} {\bibfnamefont {Xin}\ \bibnamefont {Lu}}, \bibinfo
  {author} {\bibfnamefont {Hanoh}\ \bibnamefont {Lee}}, \bibinfo {author}
  {\bibfnamefont {Michael}\ \bibnamefont {Smidman}}, \bibinfo {author}
  {\bibfnamefont {Frank}\ \bibnamefont {Steglich}}, \bibinfo {author}
  {\bibfnamefont {Piers}\ \bibnamefont {Coleman}}, \ and\ \bibinfo {author}
  {\bibfnamefont {Huiqiu}\ \bibnamefont {Yuan}},\ }\bibfield  {title} {\enquote
  {\bibinfo {title} {Strange-metal behaviour in a pure ferromagnetic kondo
  lattice},}\ }\href {\doibase 10.1038/s41586-020-2052-z} {\bibfield  {journal}
  {\bibinfo  {journal} {Nature}\ }\textbf {\bibinfo {volume} {579}},\ \bibinfo
  {pages} {51--55} (\bibinfo {year} {2020})}\BibitemShut {NoStop}%
\bibitem [{\citenamefont {Yuan}\ \emph {et~al.}(2003)\citenamefont {Yuan},
  \citenamefont {Grosche}, \citenamefont {Deppe}, \citenamefont {Geibel},
  \citenamefont {Sparn},\ and\ \citenamefont {Steglich}}]{Yuan2003}%
  \BibitemOpen
  \bibfield  {author} {\bibinfo {author} {\bibfnamefont {H.~Q.}\ \bibnamefont
  {Yuan}}, \bibinfo {author} {\bibfnamefont {F.~M.}\ \bibnamefont {Grosche}},
  \bibinfo {author} {\bibfnamefont {M.}~\bibnamefont {Deppe}}, \bibinfo
  {author} {\bibfnamefont {C.}~\bibnamefont {Geibel}}, \bibinfo {author}
  {\bibfnamefont {G.}~\bibnamefont {Sparn}}, \ and\ \bibinfo {author}
  {\bibfnamefont {F.}~\bibnamefont {Steglich}},\ }\bibfield  {title} {\enquote
  {\bibinfo {title} {Observation of two distinct superconducting phases in
  {CeCu$_{2}$Si$_{2}$}},}\ }\href {\doibase 10.1126/science.1091648} {\bibfield
   {journal} {\bibinfo  {journal} {Science}\ }\textbf {\bibinfo {volume}
  {302}},\ \bibinfo {pages} {2104--2107} (\bibinfo {year} {2003})}\BibitemShut
  {NoStop}%
\bibitem [{\citenamefont {Yuan}\ \emph {et~al.}(2006)\citenamefont {Yuan},
  \citenamefont {Grosche}, \citenamefont {Deppe}, \citenamefont {Sparn},
  \citenamefont {Geibel},\ and\ \citenamefont {Steglich}}]{Yuan2006}%
  \BibitemOpen
  \bibfield  {author} {\bibinfo {author} {\bibfnamefont {H.~Q.}\ \bibnamefont
  {Yuan}}, \bibinfo {author} {\bibfnamefont {F.~M.}\ \bibnamefont {Grosche}},
  \bibinfo {author} {\bibfnamefont {M.}~\bibnamefont {Deppe}}, \bibinfo
  {author} {\bibfnamefont {G.}~\bibnamefont {Sparn}}, \bibinfo {author}
  {\bibfnamefont {C.}~\bibnamefont {Geibel}}, \ and\ \bibinfo {author}
  {\bibfnamefont {F.}~\bibnamefont {Steglich}},\ }\bibfield  {title} {\enquote
  {\bibinfo {title} {Non-fermi liquid states in the pressurized
  ${\mathrm{cecu}}_{2}({\mathrm{si}}_{1\ensuremath{-}x}{\mathrm{ge}}_{x}{)}_{2}$
  system: Two critical points},}\ }\href {\doibase
  10.1103/PhysRevLett.96.047008} {\bibfield  {journal} {\bibinfo  {journal}
  {Phys. Rev. Lett.}\ }\textbf {\bibinfo {volume} {96}},\ \bibinfo {pages}
  {047008} (\bibinfo {year} {2006})}\BibitemShut {NoStop}%
\bibitem [{\citenamefont {Hamlin}(2015)}]{Hamlin2015}%
  \BibitemOpen
  \bibfield  {author} {\bibinfo {author} {\bibfnamefont {J.J.}\ \bibnamefont
  {Hamlin}},\ }\bibfield  {title} {\enquote {\bibinfo {title}
  {Superconductivity in the metallic elements at high pressures},}\ }\href
  {\doibase 10.1016/j.physc.2015.02.032} {\bibfield  {journal} {\bibinfo
  {journal} {Physica C: Superconductivity and its Applications}\ }\textbf
  {\bibinfo {volume} {514}},\ \bibinfo {pages} {59--76} (\bibinfo {year}
  {2015})}\BibitemShut {NoStop}%
\bibitem [{\citenamefont {Zhang}\ \emph {et~al.}(2022)\citenamefont {Zhang},
  \citenamefont {He}, \citenamefont {Liu}, \citenamefont {Li}, \citenamefont
  {Lu}, \citenamefont {Zhang}, \citenamefont {Feng}, \citenamefont {Wang},
  \citenamefont {Peng}, \citenamefont {Long}, \citenamefont {Yu}, \citenamefont
  {Wang}, \citenamefont {Prakapenka}, \citenamefont {Chariton}, \citenamefont
  {Li}, \citenamefont {Liu}, \citenamefont {Chen},\ and\ \citenamefont
  {Jin}}]{Zhang2022}%
  \BibitemOpen
  \bibfield  {author} {\bibinfo {author} {\bibfnamefont {Changling}\
  \bibnamefont {Zhang}}, \bibinfo {author} {\bibfnamefont {Xin}\ \bibnamefont
  {He}}, \bibinfo {author} {\bibfnamefont {Chang}\ \bibnamefont {Liu}},
  \bibinfo {author} {\bibfnamefont {Zhiwen}\ \bibnamefont {Li}}, \bibinfo
  {author} {\bibfnamefont {Ke}~\bibnamefont {Lu}}, \bibinfo {author}
  {\bibfnamefont {Sijia}\ \bibnamefont {Zhang}}, \bibinfo {author}
  {\bibfnamefont {Shaomin}\ \bibnamefont {Feng}}, \bibinfo {author}
  {\bibfnamefont {Xiancheng}\ \bibnamefont {Wang}}, \bibinfo {author}
  {\bibfnamefont {Yi}~\bibnamefont {Peng}}, \bibinfo {author} {\bibfnamefont
  {Youwen}\ \bibnamefont {Long}}, \bibinfo {author} {\bibfnamefont {Richeng}\
  \bibnamefont {Yu}}, \bibinfo {author} {\bibfnamefont {Luhong}\ \bibnamefont
  {Wang}}, \bibinfo {author} {\bibfnamefont {Vitali}\ \bibnamefont
  {Prakapenka}}, \bibinfo {author} {\bibfnamefont {Stella}\ \bibnamefont
  {Chariton}}, \bibinfo {author} {\bibfnamefont {Quan}\ \bibnamefont {Li}},
  \bibinfo {author} {\bibfnamefont {Haozhe}\ \bibnamefont {Liu}}, \bibinfo
  {author} {\bibfnamefont {Changfeng}\ \bibnamefont {Chen}}, \ and\ \bibinfo
  {author} {\bibfnamefont {Changqing}\ \bibnamefont {Jin}},\ }\bibfield
  {title} {\enquote {\bibinfo {title} {Record high ${T}_{c}$ element
  superconductivity achieved in titanium},}\ }\href {\doibase
  10.1038/s41467-022-33077-3} {\bibfield  {journal} {\bibinfo  {journal}
  {Nature Communications}\ }\textbf {\bibinfo {volume} {13}} (\bibinfo {year}
  {2022}),\ 10.1038/s41467-022-33077-3}\BibitemShut {NoStop}%
\bibitem [{\citenamefont {Ying}\ \emph {et~al.}(2023)\citenamefont {Ying},
  \citenamefont {Liu}, \citenamefont {Lu}, \citenamefont {Wen}, \citenamefont
  {Gui}, \citenamefont {Zhang}, \citenamefont {Wang}, \citenamefont {Sun},\
  and\ \citenamefont {Chen}}]{chenSC}%
  \BibitemOpen
  \bibfield  {author} {\bibinfo {author} {\bibfnamefont {Jianjun}\ \bibnamefont
  {Ying}}, \bibinfo {author} {\bibfnamefont {Shiqiu}\ \bibnamefont {Liu}},
  \bibinfo {author} {\bibfnamefont {Qing}\ \bibnamefont {Lu}}, \bibinfo
  {author} {\bibfnamefont {Xikai}\ \bibnamefont {Wen}}, \bibinfo {author}
  {\bibfnamefont {Zhigang}\ \bibnamefont {Gui}}, \bibinfo {author}
  {\bibfnamefont {Yuqing}\ \bibnamefont {Zhang}}, \bibinfo {author}
  {\bibfnamefont {Xiaomeng}\ \bibnamefont {Wang}}, \bibinfo {author}
  {\bibfnamefont {Jian}\ \bibnamefont {Sun}}, \ and\ \bibinfo {author}
  {\bibfnamefont {Xianhui}\ \bibnamefont {Chen}},\ }\bibfield  {title}
  {\enquote {\bibinfo {title} {Record high 36 k transition temperature to the
  superconducting state of elemental scandium at a pressure of 260 gpa},}\
  }\href {\doibase 10.1103/PhysRevLett.130.256002} {\bibfield  {journal}
  {\bibinfo  {journal} {Phys. Rev. Lett.}\ }\textbf {\bibinfo {volume} {130}},\
  \bibinfo {pages} {256002} (\bibinfo {year} {2023})}\BibitemShut {NoStop}%
\bibitem [{\citenamefont {Tomita}\ \emph {et~al.}(2001)\citenamefont {Tomita},
  \citenamefont {Hamlin}, \citenamefont {Schilling}, \citenamefont {Hinks},\
  and\ \citenamefont {Jorgensen}}]{MgB2}%
  \BibitemOpen
  \bibfield  {author} {\bibinfo {author} {\bibfnamefont {T.}~\bibnamefont
  {Tomita}}, \bibinfo {author} {\bibfnamefont {J.~J.}\ \bibnamefont {Hamlin}},
  \bibinfo {author} {\bibfnamefont {J.~S.}\ \bibnamefont {Schilling}}, \bibinfo
  {author} {\bibfnamefont {D.~G.}\ \bibnamefont {Hinks}}, \ and\ \bibinfo
  {author} {\bibfnamefont {J.~D.}\ \bibnamefont {Jorgensen}},\ }\bibfield
  {title} {\enquote {\bibinfo {title} {Dependence of ${T}_{c}$ on hydrostatic
  pressure in superconducting {MgB$_{2}$}},}\ }\href {\doibase
  10.1103/PhysRevB.64.092505} {\bibfield  {journal} {\bibinfo  {journal} {Phys.
  Rev. B}\ }\textbf {\bibinfo {volume} {64}},\ \bibinfo {pages} {092505}
  (\bibinfo {year} {2001})}\BibitemShut {NoStop}%
\end{thebibliography}%

\end{document}